\documentclass[num-refs]{wiley-article}

\usepackage{siunitx}

\papertype{Original Article}
\paperfield{Journal Section}

\title{Strong and Broadband Pure Optical Activity in 3D Printed THz Chiral Metamaterials}

\author[1,2\authfn{1}]{Ioannis Katsantonis}
\author[1]{Maria Manousidaki}
\author[1,3]{Anastasios D. Koulouklidis}
\author[1]{Christina Daskalaki}
\author[1,4]{Ioannis Spanos}
\author[1,3]{Constantinos Kerantzopoulos}
\author[1,5]{Anna C. Tasolamprou}
\author[1,6]{Costas M. Soukoulis}
\author[1,3]{Eleftherios N. Economou}
\author[1,2]{Stelios Tzortzakis}
\author[1,2\authfn{2}]{Maria Farsari}
\author[1,2\authfn{3}]{Maria Kafesaki}

\affil[1]{Institute of Electronic Structure and
Laser, Foundation for Research and Technology-Hellas, 70013
Heraklion, Crete, Greece}
\affil[2]{Department of Materials Science
and Technology, University of Crete, 70013 Heraklion,
Crete, Greece}
\affil[3]{Department of Physics,
University of Crete, 70013 Heraklion, Crete, Greece}
\affil[4]{
Department of Engineering Science,
University of Oxford, Oxford OX1 4BH, UK}
\affil[5]{
Section of Electronic Physics and Systems, Department of Physics,
National and Kapodistrian University of Athens, 15784, Athens, Greece}
\affil[6]{Ames Laboratory-U.S. DOE and
Department of Physics and Astronomy, Iowa State
University, Ames, Iowa 50011, United States}

\corraddress{I. Katsantonis, M. Farsari, M. Kafesaki, Institute of Electronic Structure and
Laser, Foundation for Research and Technology-Hellas, 70013
Heraklion, Crete, Greece}
\corremail{\authfn{1}katsantonis@iesl.forth.gr;\\\authfn{2}mfarsari@iesl.forth.gr;\\ \authfn{3}kafesaki@iesl.forth.gr}

\runningauthor{Author One et al.}

\begin{document}

\begin{frontmatter}
\maketitle

\begin{abstract}
Optical activity (polarization rotation of light) is one of the most desired features of chiral media, as it is important for many  polarization related
applications. However, in the THz region, chiral media with strong optical
activity are not available in nature. Here, we study theoretically, and 
experimentally a chiral metamaterial structure composed of pairs of vertical U-shape resonators of “twisted” arms,
and we reveal that it demonstrates large pure
optical activity (i.e. optical activity associated with negligible transmitted wave ellipticity)
in the low THz regime. The experimental data show
polarization rotation  up to $25^\circ$ for an unmatched bandwidth
of 1 THz (relative bandwidth $80\%$), from a 130 $\mu m$-thickness structure, while theoretical
optimizations show that the rotation can reach $45^\circ$. 
The enhanced chiral response of the structure is analyzed through an 
 equivalent RLC circuit model, which provides also simple optimization rules for the enhancement of its chiral response.
 The proposed chiral
structures allow easy fabrication via direct laser writing and electroless 
metal plating, making them suitable candidates for polarization control
applications. 

\end{abstract}
\end{frontmatter}

\section{Introduction}
Terahertz science is an active field of research both for its theoretical aspects
but also due to the many associated applications
\cite{Lee201465,Lee20141057,Tasolamprou2019720,Goulain20211097,
Koulouklidis20223075,Degl'innocenti20221485}, especially in sensing, imaging and
future communications. The exploitation of THz waves in those applications,
besides efficient sources and detectors, requires high-performance optical components,
such as modulators, wave-plates, lenses, etc.  Hence, devices that can manipulate
the amplitude and the phase of terahertz waves in an efficient and flexible way
have gathered great attention \cite{Padilla2006,Manceau2010,Shen2011}. Many
innovating THz devices \cite{Stoik2010106} and applications
\cite{Nagatsuma2016371,Fan2014} require  polarization control, such as
polarization rotation. Proposed ways of
realizing polarization rotators are based mainly on non-chiral anisotropic
metasurfaces \cite{Shaltout20144426,Cong2012,Zhao2011} or multilayer wire-grid
polarizers  oriented in different directions \cite{Cong2013,Sarsen2019}.
However, a crucial drawback of these approaches is that they work only for one
polarization of the incident wave. In order to realize reciprocal polarization
rotating devices insensitive to the polarization of the incident wave (at least
for normal incidence) a chiral structure is necessary
\cite{Niemi20133102,Meskers20222324,Galiffi2022}. Chiral metamaterial structures,
i.e. metamaterials with unit-cells lacking any mirror-symmetry plane, have
been shown in recent years able to give, among other effects, strong
polarization rotation (often called optical activity) even from ultrathin structures
\cite{Decker20101593,Plum2009,Hannam2014,Oh2015,Luo2017,Park2019}, providing 
high efficiency which is limited only by dissipation loss and design optimization.  
We have to stress here that not only optical rotation but all the  
metamaterials-originated strong chiro-optical effects,  e.g. 
circular
dichroism \cite{Oh2015,Caloz202058,Li2010} and asymmetric transmission \cite{Menzel2010,
Mutlu2012,Pfeiffer2014,Katsantonis2020}, have been shown to be instrumental for many
applications requiring wave polarization control 
\cite{Gansel20091513,Decker20101593,Wang2009,Kenanakis201412149}, including
sensing \cite{Tang2010,Graf2019482,Schäferling2012,Mohammadi20182669,Kilic2021,Katsantonis2022}
and spectroscopy \cite{Berova2012,Sofikitis201476}. These features may also be
reconfigurable if combined with a THz tunable material like graphene \cite{Kim2017,ZHANG202124804}. 

The key point in the metamaterial-based chiro-optical devices is that they exhibit
chiro-optical effects  orders of magnitude larger than natural
chiral media; this is due to their macroscopic nature, where the currents 
responsible for chirality are not restricted by the atomic size. Furthermore,
the scalability of  metamaterials allows for operation in almost all frequency
spectrum, provided that the available technology allows for their physical
implementation. Indeed, chiral metamaterials based on bilayer-metal chiral meta-atoms \cite{Oh2015,Park2019},
such as pairs of crosses \cite{Zhao201014553} or gamadions \cite{Wang2009}, 
have demonstrated (with proper scaling) large optical
activity in frequencies ranging from microwaves to
the optical range. 
This type of chiral metamaterials supports multiple low-frequency resonant modes dominated
by either an electric dipole resonant response or a magnetic dipole response, 
accompanied with  chirality resonance; thus  all chiro-optical effects are maximized
at the resonances. Resonances though are associated also with high wave absorption and
large impedance mismatch with the surrounding medium. Thus, the high optical 
rotation at resonance is always
accompanied by low transmittance and by high circular-dichroism \cite{Decker20101593,Khanikaev2016} (absorption 
difference between left-handed and right-handed circularly polarized waves), leading
to unavoidable non-negligible ellipticity of the transmitted wave, an  effect
undesirable for many applications requiring linearly polarized waves. This
high ellipticity of the bilayer metallic structures is boosted by the dielectric
spacer separating the two
metallic layers of the bilayer, where there is high electromagnetic field
concentration and thus maximization of the absorption response. Thus,
achievement of pure optical rotation (optical rotation associated with close
to zero ellipticity) in bilayer-metal and in most of the resonant chiral
structures seems possible only
in frequency bands between resonances, with usually moderate rotation values
and narrow-band response, features that inhibit the practical exploitation
of the related structures. 


In this work we propose a chiral metamaterial structure/design which shows
large and ultra-broadband pure optical activity in the low THz regime. The
structure is a metasurface made of three-dimenisional (3D)  metallic elements;
it is dielectrics-free
and possesses four-fold rotational symmetry. The building block (meta-atom)
is a pair of vertical U-shape resonators of "twisted" arms, as shown in
Figure ~\ref{fig1}. This type of meta-atoms can be easily fabricated via
Direct Laser Writing (DLW) \cite{Sakellari2017,Tsilipakos2020,Kenanakis2015287}
and subsequent selective metallization by, e.g., electroless plating. The geometry
of our chiral design allows co-linear electric dipole and magnetic dipole
moments (for normally-incident waves, i.e. along-$z$ in Figure ~\ref{fig1}),
resulting to bi-isotropic chiral response.  We present an extensive
theoretical and numerical analysis of the design and  demonstrate its
potential for strong and broad-band  optical rotation accompanied with
very low ellipticity. The numerical results are validated by corresponding
experimental data, validating also the large potential of
our structure in the control of THz wave polarization.

The paper is organized as follows: Initially we present the proposed
structure and demonstrate numerically its strong and broad-band pure optical activity response.
 Then, we describe
the fabrication procedure and present the experimental 
results that validate the theory and reveal also experimentally the enhanced performance of the structure in terms of optical activity. Further, to explain the response of the structure, we employ
a simple equivalent RLC circuit model for chiral metamaterials, which allows to derive simple optimization rules for achieving enhanced chiral response. In the
conclusion, we summarize the main results and suggest future perspectives
of this work.

\section{Chiral Structure and its calculated electromagnetic response}

\begin{figure}[!ht]
\centering
\includegraphics[width=14cm]{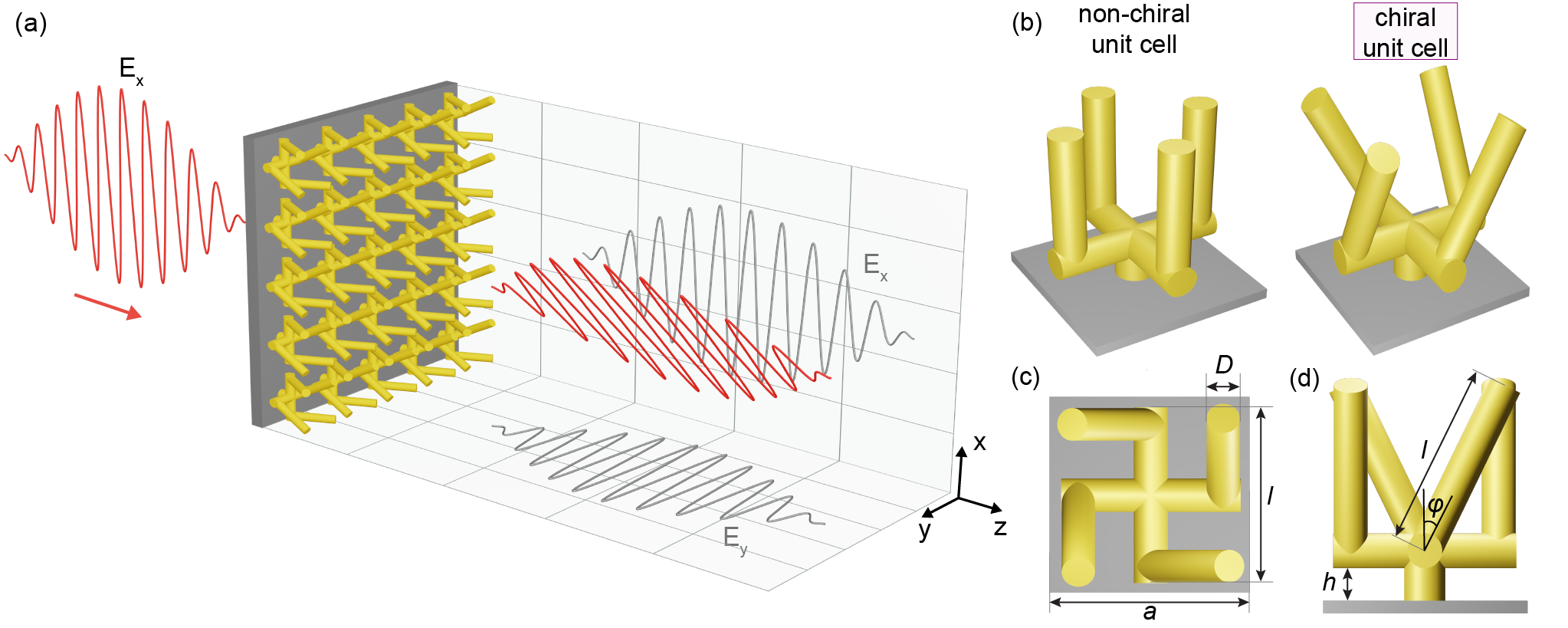}
\caption{\label{fig1}(a) Illustration of our chiral metamaterial structure. Yellow color indicates the metallic components and light-gray the silicon substrate.
        (b) Perspective view of our structure unit cell before (left panel) and after (right panel) the twist of the vertical arms. 
        (c) Top view of the chiral unit-cell and (d) side-view of the unit-cell. 
        The geometrical parameters are the following:  Lattice constant  $a=120$
        $\mu m$, arm length $l=96$ $\mu m$ (for both horizontal, i.e. at $x-y$ plane, and non-horizontal arms),  arm diameter $D=24$ $\mu m$
        and vertical support-leg height $h=18$ $\mu m$ (Note that the support-leg does not affect the EM response of the structure.).}
\end{figure}

A schematic representation of the chiral metamaterial (CMM) design proposed in
the present study is illustrated in Figure ~\ref{fig1}. It consists of a square
(in $x-y$ plane) arrangement of chiral  meta-atoms, where each meta-atom is
formed by two perpendicular metallic U-shape rings of "twisted vertical" arms; i.e., 
each initially vertical (to $x-y$ plane - see Fig. 1(b)) arm of the U-rings is rotated by an angle $\phi$
anti-clockwise (as seen from top - see Fig. 1(c)), in respect to its initial U-plane. 
The twist (rotation) of the vertical arms induces a magnetoelectric coupling in
the structure, resulting to the chiral response. In the
absence of this twist (i.e. $\phi=0$) the system behaves as typical non-chiral, split-ring-resonator-type 
metamaterial. 
The metal of the  U-shape rings constituting the meta-atoms is  Silver (Ag);
its conductivity is considered in the simulations to be linearly dependent on
frequency, ranging from $\sigma=5.0\times10^7$ S/m at 0.1 THz to
$\sigma=0.86\times10^7$ S/m at 2.2 THz \cite{Yang2015,Laman2008,Kim2013,Chen2021}.
The structure stands on a Silicon substrate (relative permittivity
$\epsilon_{Si}=11.9$ and loss tangent $\tan\delta=0.02$), while the short
(of height $h=18$ $\mu m$) vertical leg joining the rings with the substrate (it is
electromagnetically inactive) serves the complete metal-plating of the horizontal
U-arms in the fabricated  structure. The lattice periodicity is $a=120$ $\mu m$
and the length of each arm is $l=96$ $\mu m$. The arm diameter is
$D=24$ $\mu m$.

Scattering experiments/simulations provide a complete description of electromagnetic wave 
transmission and reflection by a structure. For chiral structures, where the 
eigenwaves are the circularly polarized waves, the scattering problem is 
usually formulated for circularly polarized light. However, there are many 
applications requiring linearly polarized waves while the experimental data 
taken, e.g., from the spectrometers are usually obtained for linearly polarized
electromagnetic fields. Having  the reflection and transmission coefficients 
either for circularly or from linearly polarized waves one can calculate the 
main chirality related phenomena, i.e.  optical activity and cicrular dichroism. 

To demonstrate the  broadband pure optical activity of our structure, we consider
a  unit cell as the one shown in Figure 1, with periodic boundary conditions 
along $x$ and $y$ directions, and calculate the transmitted fields for 
normally incident linearly polarized waves (using the CST Studio commercial software). In that case the incident (in) and transmitted (tr) electric fields (${\bf E}$)
are related via $T_{L}$ matrix as 
\begin{eqnarray}
\left[\begin{array}{cc}
E_{x}^{(tr)}  \\
E_{y}^{(tr)} 
\end{array}\right]=\! 
\left[\begin{array}{cc}
t_{xx} & t_{xy} \\
t_{yx} & t_{yy}
\end{array}\right]
\left[\begin{array}{cc}
E_{x}^{(in)}  \\
E_{y}^{(in)} 
\end{array}\right]= T_{L} \left[\begin{array}{cc}
E_{x}^{(in)}  \\
E_{y}^{(in)} 
\end{array}\right].
\label{eq1}
\end{eqnarray}
In Eq. (\ref{eq1}) $t_{xx}$, $t_{yy}$,$t_{xy}$ and $t_{yx}$  are the complex 
transmission coefficients, where the first 
subscript indicates the output wave
polarization and the second the incident wave polarization.  Due to the
fourfold rotational symmetry of our structure $t_{xx}=t_{yy}$ and $t_{xy}=-t_{yx}$. 

To evaluate the chiral response of our structure, we need to calculate
the transmitted wave ellipticity,  $\eta=0.5 \tan^{-1}\left[(|t_{++}|^2 - |t_{--}|^2)/(|t_{++}|^2 + |t_{--}|^2)\right]$,  directly connected to the circular dichroism,
$CD=|t_{++}|^2 - |t_{--}|^2$, as well as the polarization
rotation angle, $\theta=(1/2) [\arg{(t_{++})}-\arg{(t_{--})}]$,
a measure of the optical activity. 
To obtain $\eta$ and $\theta$ we need the
corresponding transmission coefficients for right-handed  and left-handed
circularly polarized waves, $t_{++}$ and $t_{--}$ respectively;  they can be
obtained from the corresponding linear polarization coefficients using the general formula \cite{Menzel2010}
\begin{eqnarray}
T_{CP}=\left[\begin{array}{cc}
t_{++} & t_{+-} \\
t_{-+} & t_{--}
\end{array}\right]=\frac{1}{2}\! 
\left[\begin{array}{cc}
(t_{xx}+t_{yy})+ i (t_{xy}-t_{yx})  & (t_{xx}-t_{yy})- i (t_{xy}+t_{yx}) \\
(t_{xx}-t_{yy})+ i (t_{xy}+t_{yx}) & (t_{xx}+t_{yy})- i (t_{xy}-t_{yx})
\end{array}\right].
\label{eq2}
\end{eqnarray}
which is simplified in our case taking into account the symmetries $t_{xx}=t_{yy}$ and $t_{xy}=-t_{yx}$ (these symmetries result to $t_{+-}=t_{-+}=0$).

\begin{figure}[!ht]
\centering
\includegraphics[width=14cm]{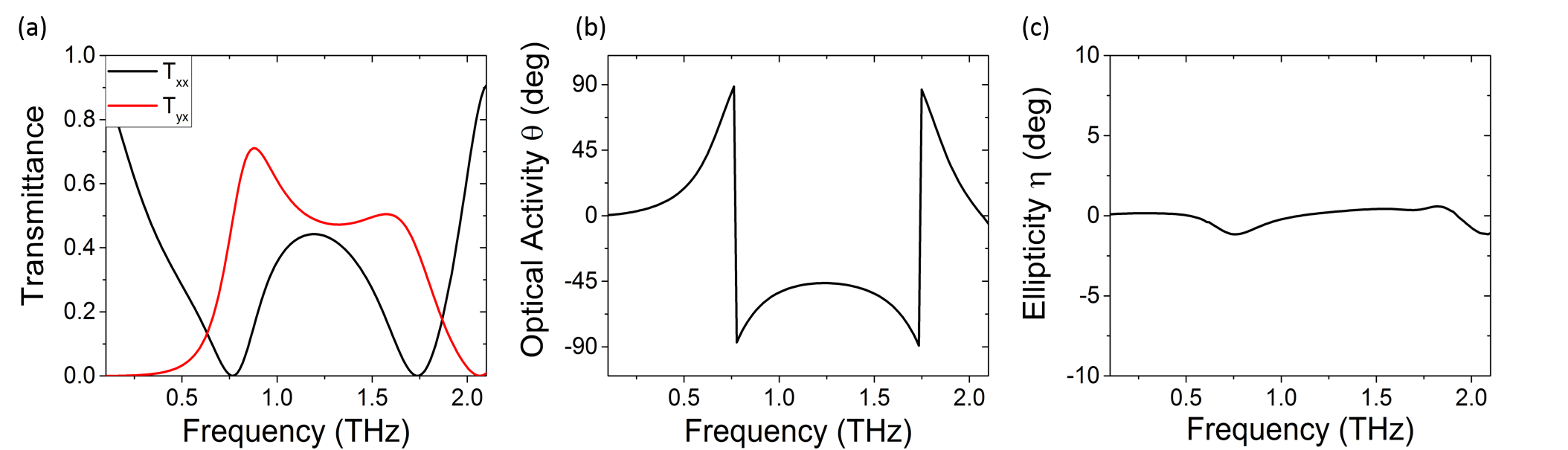}
\caption{\label{fig2} Panel (a) depicts the numerically calculated co- and
                cross-polarized transmittance spectra (black-line, $T_{xx}$, 
                and red-line, $T_{yx}$, respectively) for the structure of Fig. 1. 
                Panels (b) and (c) illustrate the corresponding optical activity and
                ellipticity of the transmitted wave.
}
\end{figure}

The calculated co- and cross-polarized transmittances
($T_{xx}=|t_{xx}|^2,T_{yx}=|t_{yx}|^2 $) as well as the corresponding optical
rotation, $\theta$, and ellipticity, $\eta$, for our structure are illustrated in 
Figure ~\ref{fig2} (a)-(c). We observe two strong  resonances, at ~0.8 THz
and ~1.8 THz, associated with dips in co-polarized transmittance and peaks in the cross-polarized one. 
In the case of no-twisted U-arms ($\phi=0$) both resonances are coming from the U-ring which is perpendicular to the incident magnetic field, as shown by field and current simulations. The first resonance is a typical magnetic resonance,
excited both by the incident magnetic field and the incident electric field (the latter due to the bianisotropy of the
ring \cite{Marqués20021444401,Katsarakis20042943}, coming from the asymmetry
of the U-shape in the direction of the electric field). The second resonance 
is predominantly the electric dipole resonance of the parallel to the electric
field U-ring; it is though   strongly affected by the coupling between
neighboring unit-cells (note that the "vertical"  U-arms of
neighboring cells form pairs of parallel cut-wires supporting also resonant
magnetic response). The twisting of the U-arms ($\phi \neq 0$) results to excitation of both U-rings forming our meta-atom;  for both rings   co-linear electric and magnetic dipoles are excited,
resulting to chiral response and thus to the resonant cross-polarized transmission
 shown in Figure 2(a).

Using the transmission data for linearly polarized waves, we obtain the corresponding ones for cicrulary polarized waves (via Eq. (~\ref{eq2})) and through them we evaluate the
optical activity and transmitted wave ellipticity for our structure. The corresponding results are shown in Figures. 2(b) and 2(c);
they reveal a quite impressive pure optical activity (i.e. optical activity associated with negligible ellipticty). The optical activity is larger than $45^{\circ}$ in the range from ~0.75 to 1.75 THz, i.e. in a relative bandwidth $\Delta f \simeq 80\%$ ($\Delta f=(f_{max}-f_{min})/(f_{max}+f_{min})/2$). The ellipticity in this range is always lower than $2^{\circ}$. This small ellipticity is attributed partially to the dielectrics-free nature of the structure. 

Thus, we can
summarize that for our proposed design, numerical simulations demonstrate   large, pure and broadband optical
rotation of  linearly polarized waves. This enhanced performance will be verified by the associated experimental studies presented below, and will be discussed further and analyzed in Section 4.


\section{Fabrication and electromagnetic characterization}
\begin{figure}[!ht]
\centering
\includegraphics[width=13cm]{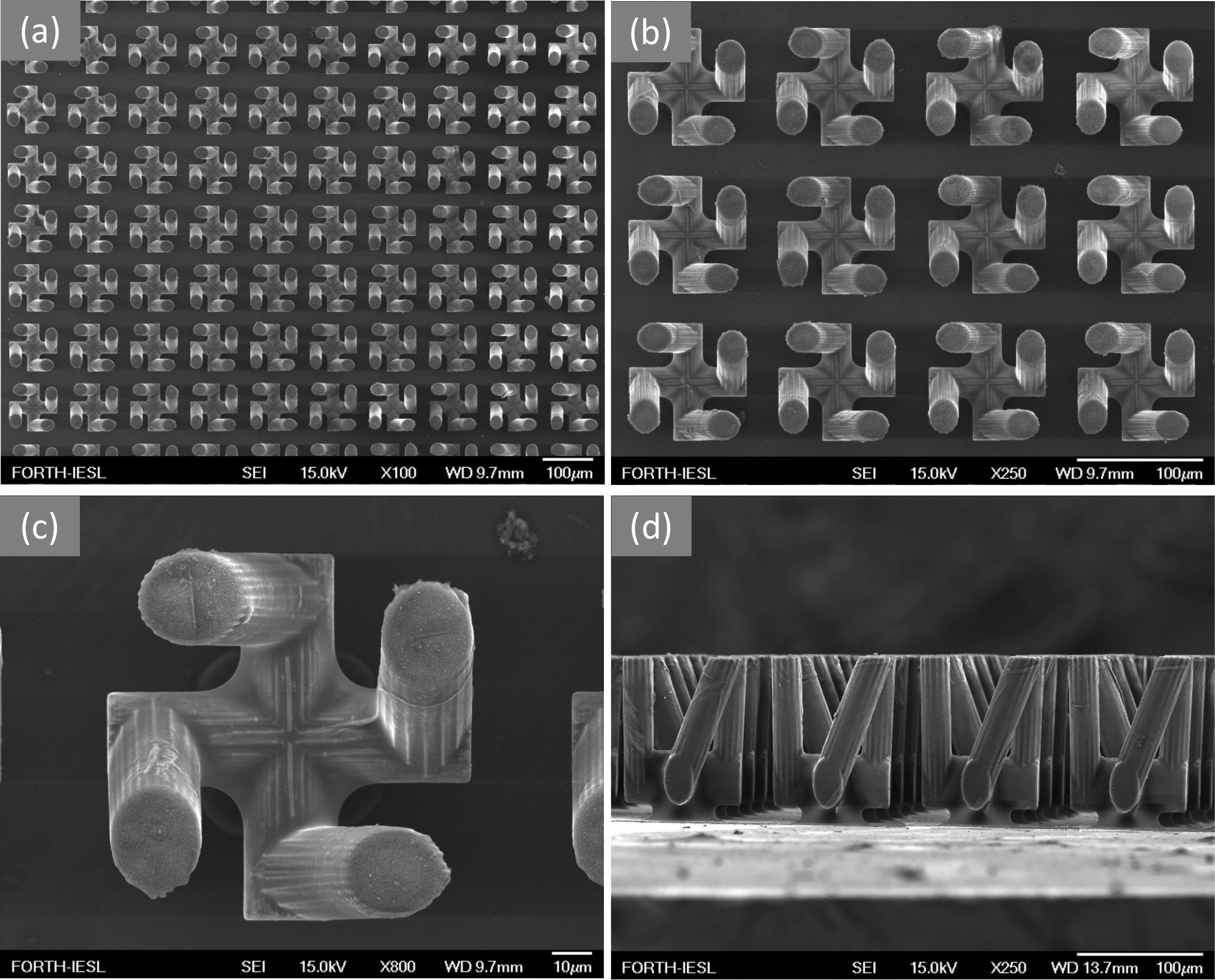}
\caption{\label{fig3} Scanning Electron Images (SEM) of the metalized Chiral THz
            Metamaterial structures fabricated on a silicon substrate. 
            (a, b, c) Top view, (d) Side view. The geometrical dimensions of 
            the structure are measured to be: 
            lattice periodicity $a=121.4$ $\mu m$, arm length $l=92.8$ $\mu m$,
            arm diameter  $D=24$ $\mu m$ and angle of the arm  $\phi=21^{\circ}$.}
\end{figure}

In order to fabricate the metamaterial structure depicted in
Figure ~\ref{fig1}, we employed Direct Laser
Writing (DLW) by Multiphoton Polymerization (MPP)
\cite{Ovsianikov2008,MALINAUSKAS20131}, followed by selective metallization
through electroless plating (EP) with silver \cite{Aristov2016}. 
The detailed experimental setup used for the fabrication and
the subsequent metallization of the fabricated structures is presented 
at the Experimental Section of this paper. 
Scanning Electron Microscope Images (SEM) of
the fabricated metalized chiral THz metamaterial structures are
shown in Figure ~\ref{fig3} in top and side view.
The geometrical parameters of the metallic structures are measured to
be the following: the lattice constant  $a=121.4$ $\mu m$,  the arm length
$l=92.8$ $\mu m$, the arm diameter  $D=24$ $\mu m$ and the angle of the arm  $
\phi=21^{\circ}$ (note the deviations in the geometrical parameters from the optimized simulated structure of Figure 2).

\begin{figure}[!ht]
\centering
\includegraphics[width=14cm]{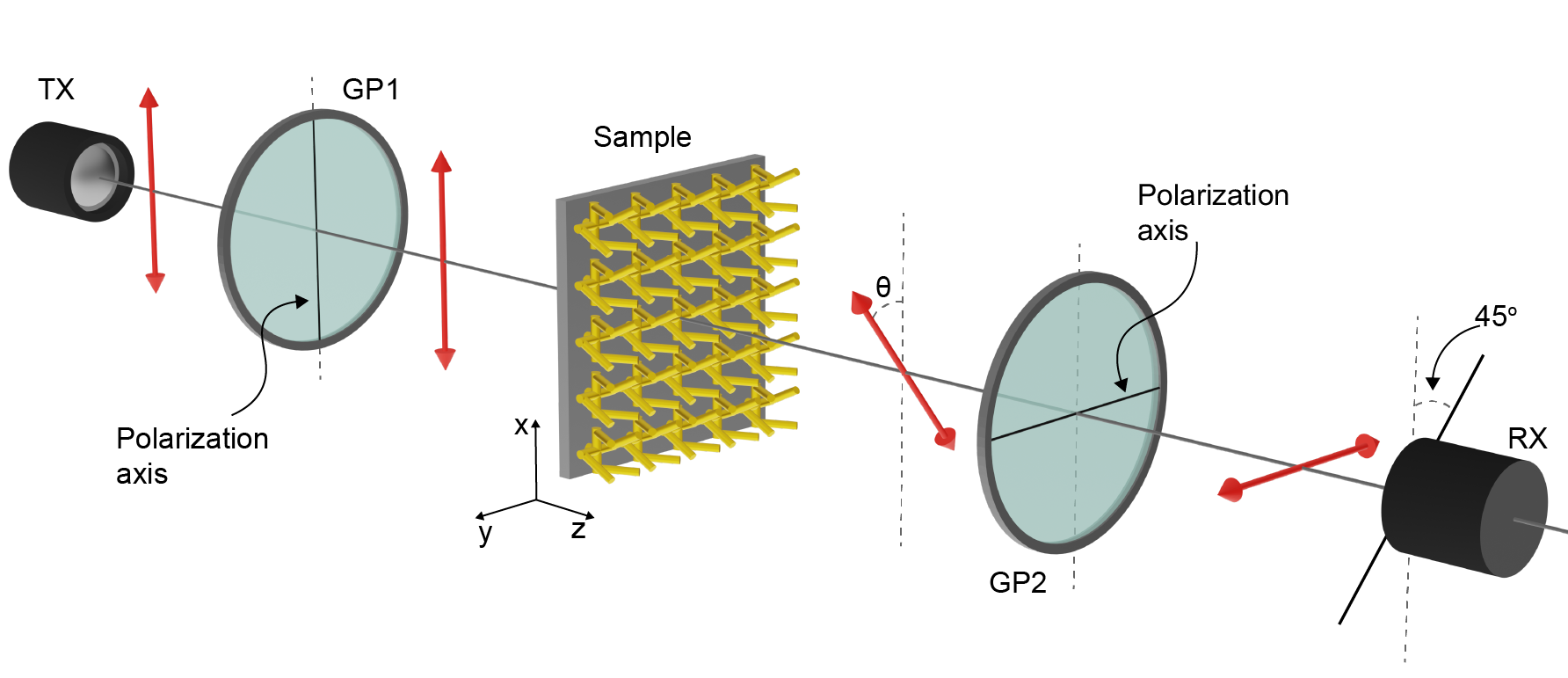}
\caption{\label{fig:setup}
        Schematic representation of the experimental setup. TX:THz emitter, 
        GP1,2:Wire grid polarizers, RX: THz receiver, rotated at $45^{\circ}$
        with respect to x-axis.}
\end{figure}

\begin{figure}[!ht]
\centering
\includegraphics[width=11cm]{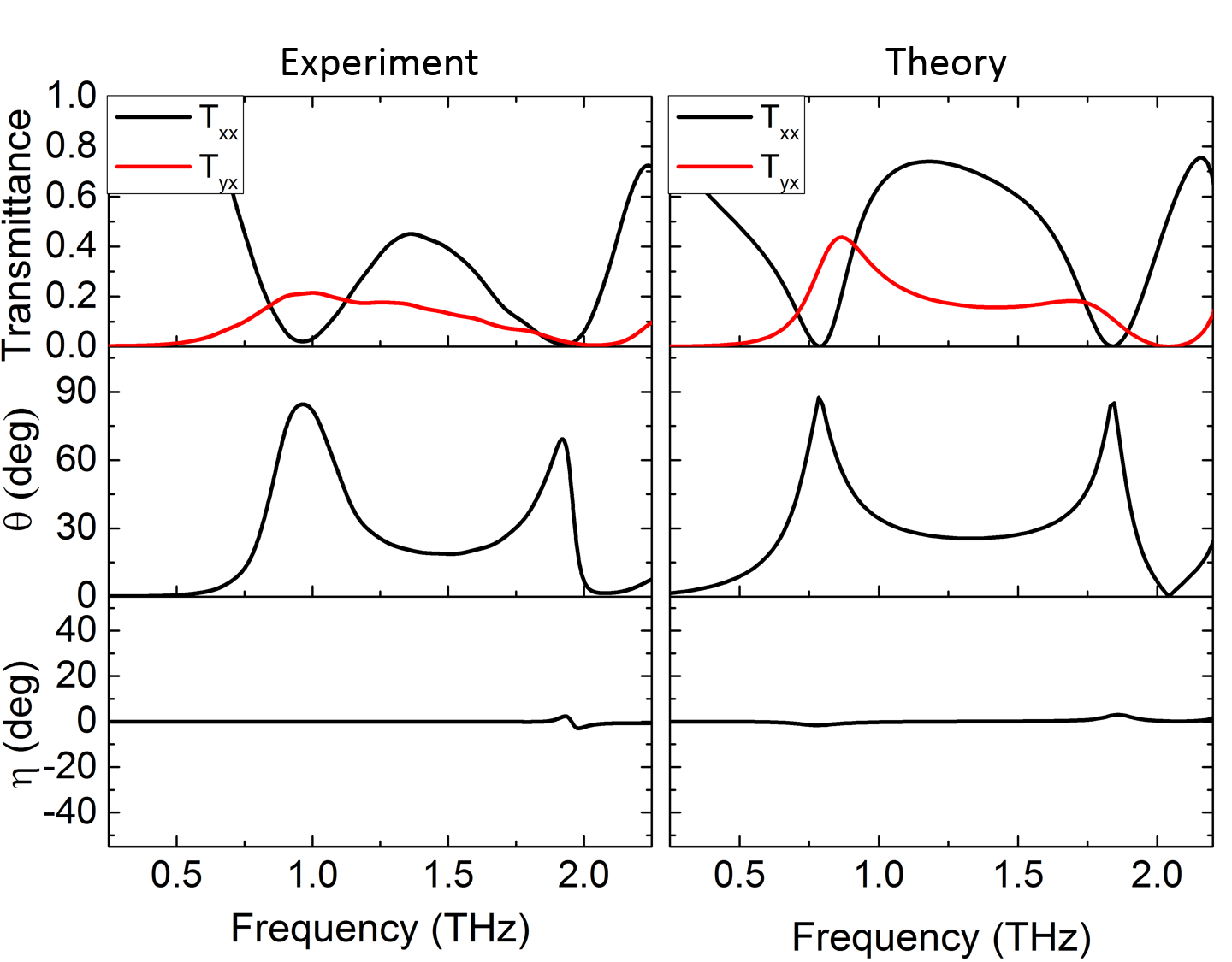}
    \caption{\label{fig5} The three left-hand side panels show experimentally measured  co- and cross- polarized transmittances (top left),
    optical activity, $\theta$ (middle left), and ellipticity, 
    $\eta$ (bottom left). The corresponding simulation results are illustrated
    in the right-panels and are in good agreement with the experimental data.}
\end{figure}

To determine the optical characteristics of our chiral metamaterial, we used a
terahertz time-domain spectroscopy (THz-TDS) setup based on photoconductive
antennas (TOPTICA TeraFlash pro) operating in transmission mode. A schematic
representation of the experimental setup in shown in Figure ~\ref{fig:setup}. A
broadband THz pulse, linearly polarized along the x-axis, is generated by the
photoconductive emitter (TX). The THz pulse passes through a wire grid polarizer
(GP1) that further defines the polarization along the same axis and impinges on
the sample at normal incidence. The transmitted by the sample wave, then, passes
through a second wire grid polarizer (GP2) which can be either parallel or
perpendicular to GP1. This way, cross- or co-polarized transmission measurements
can be performed. The THz wave is finally guided to the photoconductive detector
(RX). Since RX is highly sensitive to linearly polarized THz waves, it was
rotated by $45^{\circ}$ with respect to x-axis. This ensures that there is
always an equal component of the THz field along 
the x- and y-axis~\cite{Li201725842,Yang2020}. 
The THz transmission spectra were normalized to a co-polarized transmission spectrum
through a bare silicon substrate. Figure ~\ref{fig5} shows the measured transmission power spectra ($T_{xx}=|t_{xx}|^2$ and
$T_{yx}=|t_{yx}|^2$) of our metamaterial at linear polarization incidence, well as the corresponding optical activity, $\theta$, and ellipticity $\eta$.

 To compare our experimental results to the numerical ones, the same calculations
 presented before are used. The corresponding data are shown also in Figure 5, next to the experimental ones. For the  calculations We assume the following geometrical parameters which are
 the same as the fabricated sample (with a deviation less than 1 $\%$) with bulk 
 Ag conductivity in the range from $\sigma=8.7\times10^6$ S/m at 0.1 THz to 
 $\sigma=3.5\times10^6$ S/m at 2.2 THz as in \cite{Aristov2016}. Particularly,
 the lattice constant is $a=120$ $\mu m$, the arm length is $l=90$ $\mu m$ 
 while each arm is rotated by an angle $\phi=20^{\circ}$, the arm diameter 
 is $D=24$ $\mu m$ and the support vertical leg is $h=18$ $\mu m$. As expected,
 strong polarization rotation with almost zero ellipticity is obtained, for a
 bandwidth of 1 THz.
The optical measurements of our fabricated sample represent this trend very 
nicely, as can be observed in the left column of Figure~\ref{fig5}. In particular,
 the measured co-polarization 
transmission amplitude,$T_{xx}$ in the spectra range from 1 to 2 THz is compressed below
50 $\%$. At the same time, the cross-polarization transmission amplitude, $T_{yx}$,
reaches its maximum value of 20$\%$, indicating effectively the polarization
rotation of linearly polarized waves. Using the experimental data, we evaluate the
polarization rotation angle and ellipticity as depicted in the left-hand side
panels of Figure~\ref{fig5}. We observe optical activity up to $25^{\circ}$ for
a frequency range over 1 THz accompanied by zero ellipticity. However, the
experimental results are a little weaker as well, as the first resonance appears
at somewhat higher frequency than expected from simulations. This difference
can be attributed mainly to the difference in the conductivity of
the Silver coated sample compared to the  bulk
conductivity used in the simulations as well as to fabrication imperfections,
especially in the cross-sections of the cylindrical arms which are in some cases elliptical instead of circular. Furthermore, at low THz  the skin-depth  of  Silver,
strongly depends on the value of the conductivity \cite{Zhou202236712}, and
in some cases the skin depth can be larger than the achieved of the deposited metal.

\section{Theoretical analysis and discussion}
\begin{figure}[!ht]
\centering
\includegraphics[width=14cm]{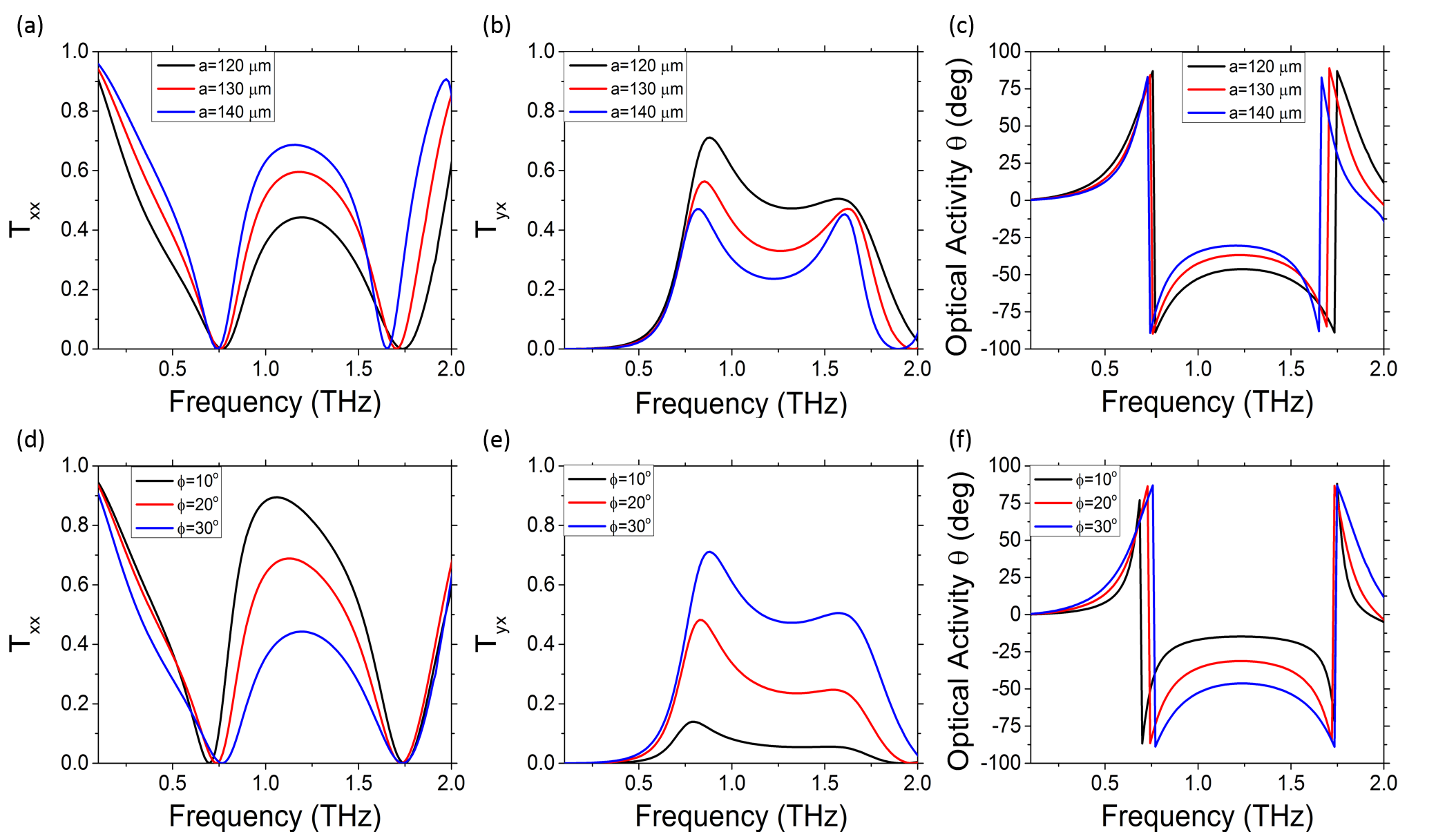}
\caption{\label{fig6} The top panels (a)-(c) depict the numerically calculated
        co- and cross- transmittance spectra and the corresponding optical activities
        at different lattice-constants ($a$) of our metamaterial. In the bottom panels (d)-(f),
        the numerically calculated co- and cross-polarized  transmittance spectra and the
        corresponding optical activities at three different rotation angles of the U-rings arms: $\phi=10^o$,
        $\phi=20^o$ and $\phi=30^o$, of arms.}
\end{figure}

 In this section we discuss first the sensitivity 
 of the electromagnetic
 response of our design for different geometrical parameters. In particular, we demonstrate the
 effect of changing the in-plane lattice constant (affecting the coupling with the nearest neighbors) and the vertical arms rotation angle ($\phi$)
 on the transmittance features  as a function of frequency. 
 The co- and cross- polarized transmittances
($T_{xx}=|t_{xx}|^2,T_{yx}=|t_{yx}|^2 $) for different lattice-constant values ($a$) as well as the corresponding optical 
 rotations, $\theta$, are illustrated in Figure ~\ref{fig6} (a)-(c). We observe
 that the increase of the lattice constant leads to lower cross-polarized transmittance and optical rotation values. It does not affect strongly though the position of the resonances, indicating that the nearest-neighbor coupling is not a parameter determining or dominating the structure response.
 
 Regarding the effect of the rotation angle, $\phi$, it is demonstrated in Figs. ~\ref{fig6} (d)-(f). 
 Since  the system is non-chiral in the absence of
 rotation, $\phi=0$, of the arms, we expect that by tuning the angle $\phi$ we
 can highly control the optical activity. Indeed, the numerical data of Figs. ~\ref{fig6} (d)-(f)
 verify this tunability potential, demonstrating once again the 
  strong, broadband 
 and relatively flat pure optical activity of our structure. 
 

To understand and explain the calculated and measured response of our structure, we perform a theoretical analysis for obtaining    qualitative formulas for the material parameters determining the structure response, and in particular for  evaluating the effective chirality close to the first structure resonance.

We consider our meta-atom element consisting of two perpendicular U-shaped rings, as shown in Figure 7, and we examine the response of both rings to an incident EM field. 
Each U-ring is considered as a polarizable particle showing electric and magnetic dipole response; the electric and magnetic dipole moments, {\bf p} and {\bf m} respectively, are connected with the local  fields  by \cite{Niemi20133102,Asadchy20181069}
\begin{equation} \label{eqthree}
\textbf{p}=\Bar{\Bar{\alpha}}_{ee}\textbf{E}+\Bar{\Bar{\alpha}}_{em}\textbf{H}, \hspace{5mm}
\textbf{m}=\Bar{\Bar{\alpha}}_{mm}\textbf{H}+\Bar{\Bar{\alpha}}_{me}\textbf{E}  
\end{equation}
where $\Bar{\Bar{\alpha}}_{ee}$, $\Bar{\Bar{\alpha}}_{mm}$, $\Bar{\Bar{\alpha}}_{em}$, $\Bar{\Bar{\alpha}}_{me}$
 are the electric, magnetic, electromagnetic, and magnetoelectric polarizability
 tensors of the U-ring and the  $\textbf{E},\textbf{H}$ indicate  the local 
 fields (i.e. external fields plus fields generated by the induced currents at the ring; we omit for simplicity any coupling between  unit cells, and we consider operation in the quasistatic limit).

\begin{figure}[!ht]
\centering
\includegraphics[width=11.5cm]{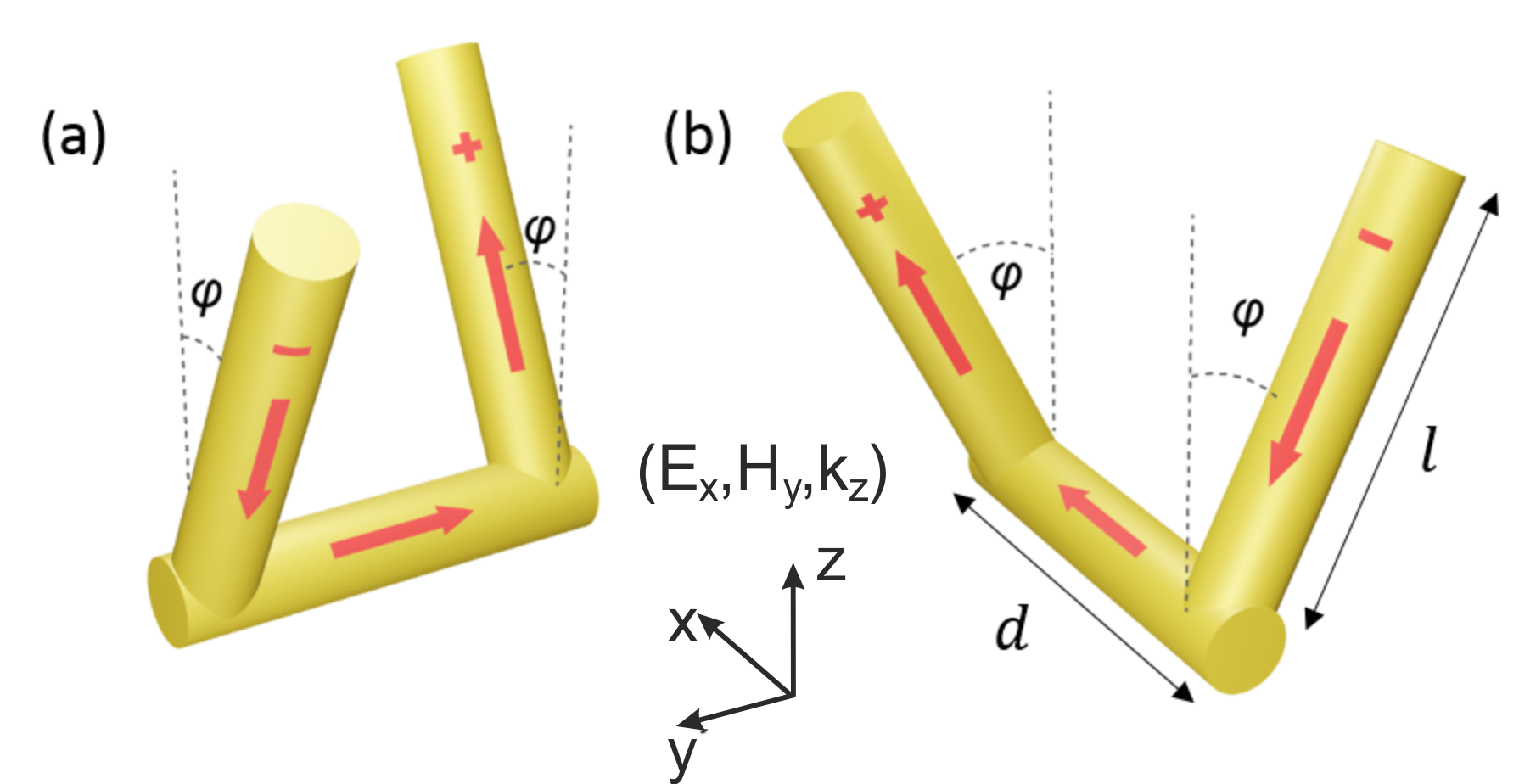}
        \caption{\label{fig7} Topologies of the two basic chiral metallic
        elements constituting the meta-atom of our structure (the meta-atom is the addition of the two elements). Both elements are the same U-shaped ring of twisted (by an angle $\phi$) vertical arms. The components of the incident electromagnetic wave are also marked in the figure, along with the currents they excite (red arrows) in the frequencies around the first structure resonance (magnetic-type resonance).}
\end{figure}

Considering the incident field configuration shown in Fig. 7, the induced currents at the lowest frequency resonance are as shown by the red arrows. Calculating the currents and/or the accumulated charges one can evaluate the electric and magnetic dipole moments relevant to each ring  and through them the individual polarizability elements.
To evaluate the currents we consider each U-ring  as an effective  RLC circuit described by the basic Kirchhoff equation  
\begin{equation} \label{eqsix}
L\frac{d I_{a,b}}{dt}+R I_{a,b}  + \frac{Q_{a,b}}{C} =U_{a,b}
\end{equation}
where the subscripts a and b refer to the configuration of Figures 7a and 7b, respectively, $I$ is the current, $Q$ the corresponding  charge ($I=dQ/dt$ ), $L$ the inductance, $R$ the resistance
and $C$ the capacitance of each configuration. The source term, $U$, is the electromotive force ($EMF=\int {\bf E \cdot} d{\bf l_c} + \mu_0 {\bf S \cdot }d{\bf H}/dt$; $l_c$: conductor length, $S$: loop-current-enclosed area) resulting from both the external and the induced fields. $U$ for the configurations a and b  can be written as 
\begin{equation} \label{eqfour}
U_{a}=-E_{y} d + E_{x} 2 l \sin{\phi} - \mu_0 l d \cos{\phi} \frac{ d H_{x}}{d t} - \mu_0 l^2 \cos{\phi} \sin{\phi} \frac{ d H_{y}}{d t}
\end{equation}
\begin{equation} \label{eqfive}
U_{b}=E_{x} d + E_{y} 2 l \sin{\phi} + \mu_0 l d \cos{\phi} \frac{ d H_{y}}{d t} - \mu_0 l^2 \cos{\phi} \sin{\phi} \frac{ d H_{x}}{d t}
\end{equation}
where $\mu_0$ is the vacuum permeability and $l,d,\phi$ are the geometrical
parameters defined in Figure \ref{fig7}.

Note that, for $\phi=0$, for the configuration a the  magnetic resonance of the structure can not be excited by the incident fields
($E_x$, $H_y$), while for the configuration b both incident electric and magnetic fields are able to excite loop currents in the U-ring, i.e. the structure is bianisotropic \cite{Marqués20021444401,Katsarakis20042943}. 

Considering harmonic time dependence of  the form of $e^{-i\omega t}$, one can calculate the current in Eq. \ref{eqsix}, the corresponding charge $Q=I/(-i\omega)$ and, through them, the electric and magnetic dipole moments, ${\bf p}=Q {\bf d_{cs}}$, ${\bf m}=I{\bf S}$ (${\bf d_{cs}}$ is the charge separation and ${\bf S}$ the area enclosed by loop currents), as a function of the fields, and the polarizabilities as defined by Eq. \ref{eqthree}. The details of the corresponding calculations are given in the Supporting Information.

For qualitative analysis we can consider the total electric and magnetic dipole moments of our double-ring meta-atom as the sum of the dipole moments of configurations a and b (any coupling between the two perpendicular rings is implicitly taken into account through the modification of induced currents in each ring).
Having at hand the total electric  and magnetic dipole moments of the unit cell, 
 the  electric polarization and the magnetic polarization
can be  calculated via $\textbf{P}=\textbf{p}/V$ and $\textbf{M}=\textbf{m}/V$,
respectively, where $V$ denotes the volume of the unit cell. 
(Note that we again ignore here, for simplicity, the coupling between unit cells.)

Then the macroscopic (average) material parameters of our structure, including the chirality, which is crucial for the understanding of its chiral response, can be obtained taking
into account the standard constitutive relations
$\textbf{D}=\epsilon_0\textbf{E}+\textbf{P}=\Bar{\Bar{\epsilon}}\textbf{E}+i(\Bar{\Bar{\kappa}}/c) \textbf{H}$ and $\textbf{B}=\mu_0 (\textbf{H}+\textbf{M})=\Bar{\Bar{\mu}} \textbf{H}-i (\Bar{\Bar{\kappa}}^{T}/c)\textbf{E}$.
Applying this procedure (see Supporting Information), the effective permittivity and permeability for the double ring result to be scalar quantities (i.e. diagonal tensors with equal elements; note that for each isolated U-ring both diagonal and off-diagonal  permittivity and permeability tensor elements appear).
The chirality parameter though, $\Bar{\Bar{\kappa}}$, has both diagonal and off-diagonal elements, demonstrating the structure bianisotropy. The diagonal elements, which are the ones involved in the cross-polarized transmission \cite{Niemi20133102,Asadchy20181069}   are given by
\begin{equation} \label{eqkxx}
\kappa_{xx}=\kappa_{yy}=\frac{ \omega c \mu_0 l^2 d \cos{\phi}\sin{\phi}}{V L[\omega^2-\omega^2_{0}+i\omega (R/L)]}
\end{equation}
where $\omega^2_0=1/LC$ is the resonance frequency of the structure. The off-diagonal  elements are given by
\begin{equation} \label{eqkxy}
\kappa_{xy}=-\kappa_{yx}=\frac{ \omega c \mu_0  l \cos{\phi} (2l^2 \sin^2{\phi}+d^2)}{V L[\omega^2-\omega^2_{0}+i\omega (R/L)]}
\end{equation}

From Eq. \ref{eqkxx} one can see that the chirality strength is larger the larger the meta-atom "filing ratio" within the unit cell, a result consistent with the observed in Fig. 6 dependence of the cross-polarized transmittance and optical activity on the unit cell size. It seems also that the length of the vertical meta-atom arms plays more pronounced role in the chirality than that of the horizontal arms (for that we consider that the inductance $L$ in Eq. (\ref{eqkxx}) is proportional to the ring area ($=ld$ for $\phi=0$). Finally, as is expected, the chirality is generated by the non-zero values of the twist angle $\phi$ and it is highly affected by $\phi$, in consistency also with the results of Fig. 6.

Regarding the off-diagonal terms of the chirality tensor, the twist angle $\phi$, as is seen by Eq. (\ref{eqkxy}), enhances the strength of these terms, i.e. it enhances the strength of the bianisotropy. 
It is not clear though whether this enhancement is the origin of the reduced ellipticity, an effect observed also in \cite{Hannam2014}.


\section{Conclusion}
We have proposed a chiral metamaterial design made of 3D metallic
elements for efficient polarization control of electromagnetic waves in a broad
frequency range. Particularly, we have demonstrated both numerically and experimentally
that our 3D metamaterial, composed of vertical U-shaped resonators of twisted arms, exhibits
strong ($>45^{\circ}$)  and ultrabroadband (relative bandwidth $80\%$) optical activity with very low ellipticity in the low THz region.
The fabrication of the structure was performed through Direct Laser Writing and subsequent electroless silver plating, while the experimental electromagnetic response was demonstrated via THz time domain spectroscopy. Finally, the structures was  studied also analytically, though an equivalent RLC circuit model, and the parameters determining its response were identified.

The large, broadband and pure optical activity of our structure equips it with unique potential for polarization control applications, particularly important in the THz region where there is still a serious lack for efficient optical components.
 At the same time, our extensive theoretical and
numerical study will be instrumental in providing guidance for further expansion
and generalization in more complicated systems, where the combination of chirality
with other special symmetries or asymmetries may open a new direction in the field of THz photonics.

\section{Methods and Experimental Section}
\subsection*{Samples Preparation}
\textbf{Photosensitive Material:} \\
The material used for the fabrication of the 3D chiral metamaterial was a
zirconium-silicon organic-inorganic hybrid composite doped with metal-binding
moieties \cite{Aristov2016}.  It was produced by the addition of
methacryloxypropyl trimethoxysilane (MAPTMS) to zirconium propoxide (ZPO, 70\%
in propanol). 2-(dimethylamino)ethyl methacrylate (DMAEMA) was also added to
provide the metal-binding moieties that enabled the selective metallization of
the dielectric structures. MAPTMS and DMAEMA were used as the organic
photopolymerizable monomers, while ZPO and the alkoxysilane groups of MAPTMS
served as the inorganic network forming moieties. 4,4-bis(diethylamino)
benzophenone (BIS) was used as a photoinitiator. The photopolymerizable material
was synthesized as described in detail in Ref. \cite{VASILANT}. The samples were
prepared by drop-casting onto a 530  $\mu$m thick silanized high-resistivity
silicon substrate, and the resultant films were dried on a hot plate at
55$^{\circ}$  for 60 min before the photopolymerization. \\
\\
\textbf{Direct Laser Writing by Multiphoton Polymerization}\\
In DLW, the beam of an ultrafast laser is tightly focused into the volume of a
transparent photopolymerizable resin. Polymerization is initiated only within the focal
volume element, viz. the voxel, where the intensity is high enough to trigger
multi-photon absorption. Scanning the laser beam inside the material, 3D
structures can be directly printed, in a layer-by-layer fashion. After the
fabrication process is completed, the sample is immersed into appropriate
solvents and the unexposed resin is dissolved to reveal the freestanding 3D
structure. A droplet of
the photosensitive material was placed  onto a 530  $\mu$m thick silanized
high-resistivity silicon substrate for the photopolymerization. 
A Femtosecond Fiber Laser
(FemtoFiber pro NIR, Toptica Photonics AG) emitting at 780 nm with a pulse
duration of 150 fs, average output power 500 mW, and a repetition rate of 80 MHz
was employed as a light source~\cite{Farsari_2010}.  
A 40× microscope objective lens (Zeiss, Plan
Apochromat, N.A. = 0.95) was used to focus the laser beam into the volume of the
photosensitive material. A Galvanometric scanner-based system (Scanlabs
HurryscanII 10, computer-controlled) was used to scan the focused laser beam
through the polymeric sample following the predefined metamaterial structure
design path.  Z-axis scanning and larger-scale x-y movements were possible with
the use of a high-precision three-axis  
linear translation stage (Physik Instrumente). The structures were fabricated in
a layer-by-layer fashion starting from the bottom (vertical leg) towards the
arms of the structure with the first layer adhering to the surface of the
silicon substrate. The scanning speed used was 4000 µm/s. The power for the
fabrication of the structures was measured, before the objective, to be 175 mW.
The live-monitoring of the fabrication process was  achieved using a CCD camera,
with appropriate imaging optics. Finally, an overall 3 x $3mm^2$  metasurface of
24×24 3D chiral meta-atoms array with a periodicity constant of 121.4 $\mu m$ on
the surface of a high resistivity silicon substrate was produced, as is shown in
Figure~\ref{fig3}.\\  
\\
\textbf{Metallization:}\\
After the fabrication of the chiral metamaterials array was completed, the
metallization process of the sample followed in order for the structures to
become conductive and  gain optical activity. 
The metallization process of the
3D chiral metamaterials was based on selective electroless silver
plating according to a modified protocol based on Ref. \cite{Aristov2016}.
This protocol has been shown to offer conductivity to the microstructures of
$\sigma=(5.71\pm3.01)\times10^6$ (S/m) \cite{VASILANT}. 
EP is a fairly simple
process that does not require any specialized equipment, and the metal
deposition can be done without using any electrical potential. 
In general, it is
characterized by the selective reduction of metal ions at the surface of a
catalytic substrate immersed into an aqueous solution of metal ions, with
continued deposition on the substrate through the catalytic action of the
deposit itself.  In detail, EP
comprises three main steps: seeding, reduction, and silver plating. \\
\underline{Seeding:}
The samples were immersed in a 0.05 mol/L AgNO3 aqueous solution at
room temperature for 38 hours. This was followed by thorough rinsing with double
distilled (d.d.) water and then left to dry at room temperature. \\
\underline{Reduction:} An aqueous sodium borohydride (NaBH4) solution 6.6 M was prepared
24hours before the immersion of the samples. The solution was very well mixed
and kept uncovered to get rid of trapped air bubbles. The samples were
subsequently dipped in the solution for 22 hours to reduce the silver ions and
form silver nanoparticles. The samples were washed thoroughly in fresh d.d.
water and left to dry. \\
\underline{Silver plating:} A 0.2 M AgNO3 aqueous solution was mixed with
5.6\% NH3 (28\% in
water) and 1.9 M glucose (C$_{6}$H$_{12}$O$_{6}$ > 98\%) as a reducing agent, at
a volumetric ratio 5:3:8. The samples were immersed in the solution for a few
minutes, before the solution turned to dark.  In the meanwhile, a fresh solution
was prepared to replace the old one. This process was repeated 5 times.
\\
After the metallization process was completed, the dielectric structure was
coated by a thin silver nanoparticles sheet.  The thickness of this metal
coating on the structures was measured from SEM images, and found to be in the
range of 80 to 130 nm. This means that the core of the structure was still
dielectric. The diversity of the thickness parameter resulted from the fact that
the metallic coat is not a smooth and bulk layer of silver on the structure
surface, but a film of sprinkled metal nanoparticles of varying diameters, that
are attached to the structures.  

\subsection*{Numerical calculations}

Numerical calculations were carried out with the commercial software package CST Microwave Studio as well as with the commercial software package COMSOL Multiphysics, employing a finite element solver in the frequency domain. A fine size of tetrahedral spatial mesh was chosen according to the COMSOL physics-controlled mesh. The transmitted and reflected coefficients were simulated for one square unit cell with the periodic boundary conditions on the x- and y- sides. The incident waves were excited on the top of the simulation domain by excitation of x- or y-polarized components sweeping the input port 1 or 2, respectively. These ports also measured the reflected x- and y- polarization waves. The detectors at the bottom simulation domain, output port 3 and 4, measured the transmitted x- and y- polarization waves.  The permittivity of the Silicon substrate was assumed to be constant over the frequency range with $\epsilon=11.9$ and $\tan(\delta)=0.02$.

\section*{Supporting Information}

Supporting Information is available from the Wiley Online Library or from the author.

\section*{Acknowledgements}
This research work was partly supported by the Hellenic Foundation for Research and
Innovation (H.F.R.I.) under the "2nd Call for H.F.R.I. Research Projects to
support Faculty members and Researchers" (Project Number: 4542), and by the European Union, projects In2Sight (FETOPEN-01-2018-2019-2020, GA:964481) and FABulous (HORIZON-CL4-2022-TWIN-TRANSITION-01-02, GA:101091644). Ms. A. Manousaki provided expert SEM support. Useful communication with Prof. Thomas Koschny is also acknowledged.

\section*{Author Contributions}
I.K, A.T. and M.K. carried out the numerical simulations. I.K. and M.K. developed the theoretical model. M.M. and I.S. fabricated the samples and took the SEM images. A.K., C.D. and K.K. carried out optical characterization experiments. C.S., E.E., S.T., M.F. and M.K. contributed to physical insight, supervised the project, guided manuscript organization and edited the manuscript. All authors contributed to the preparation of the manuscript.

\section*{Conflict of Interest}
The authors declare no conflict of interest.

\bibliography{sample}

\end{document}